 \documentclass[preprint]{aastex}

\shorttitle{Periodic and Aperiodic Variability of HerX--1}
\shortauthors{Moon \& Eikenberry}

\begin{document}


\title{Discovery of Coupling between Periodic and Aperiodic Variability 
and X-ray Quasi-periodic Oscillations from Her X--1}








\author{Dae-Sik Moon and Stephen S. Eikenberry}
\affil{Department of Astronomy, Cornell University, Ithaca, NY 14853 \\
moon,eiken@astrosun.tn.cornell.edu}


\begin{abstract}
We report the discovery of coupling between periodic and 
aperiodic variability and $\sim$12-mHz X-ray quasi-periodic oscillations (QPOs)
from the X-ray binary pulsar Her X--1
using data from the {\it Rossi X-Ray Timing Explorer}.
We found two different couplings, one during the pre-eclipse dips
and the other during the normal state of the source,
using a method which directly compares 
the low-frequency power-density spectra (PDS)
with those of the sidebands around the coherent pulse frequency.
The pre-eclipse dip lightcurves show 
significant time variation of photon counts,
and this variation appears in the PDS as both strong mHz powers
and well-developed sidebands around the coherent pulse frequency.
The linear correlation coefficients between the mHz PDS and the sideband PDS
obtained from two pre-eclipse dips data segments 
are 0.880 $\pm$ 0.003 and 0.982 $\pm$ 0.001, respectively.
This very strong coupling demonstrates that the amplitudes of the coherent
pulsations are almost exactly modulated by the aperiodic variabilities,
suggesting that both the periodic and aperiodic variabilities are related 
to time variation of obscuration of X-rays from the central pulsar
by an accretion disk during pre-eclipse dips.
We also found weak coupling during the normal state of the source, 
together with $\sim$12-mHz QPOs.
The normal state coupling seems to reconcile with the prediction that
the aperiodic variabilities from X-ray binary pulsars are due to 
time-varying accretion flows onto the pulsar's magnetic poles.
If the $\sim$12-mHz QPOs are due to global-normal
disk oscillations caused by the gravitational 
interactions between the central pulsar
and the accretion disk, the inferred inner-disk radius
is roughly comparable to the magnetospheric radius,
$\sim$1 $\times$ 10$^8$ cm.
\end{abstract}




\section{Introduction}

Timing analysis of rapid aperiodic variabilities,
including quasi-periodic oscillations (QPOs),
has been one of the most important
tools for studying various types of X-ray binary systems.
Since the aperiodic variabilities are generally thought to be related to the innermost motions
of accretion disks, the analysis of aperiodic variabilities can give us
useful information about the interaction between 
accretion disks and the central object in X-ray sources.
For example, the power-density spectra (PDS) obtained from observed lightcurves
are often used to distinguish between X-ray binary systems containing
black holes and neutron stars,
to infer rotation rates of neutron stars in low-mass X-ray binary systems,
and to investigate accretion-powered X-ray binary pulsars (AXBPs)
(e.g., van der Klis 1995).

Among different types of X-ray binaries, 
AXBPs have unique characteristics in timing analysis.
The main time-varying component of AXBPs is periodic coherent pulsations
from the central neutron star,
and the coexistence of the periodic and aperiodic variabilities makes 
timing analysis 
much more complicated than the purely aperiodic cases.
This complication often appears as PDS heavily 
contaminated by numerous bumps and wiggles 
(van der Klis 1995).
On the other hand,
detection of periodic coherent pulsations from 
an X-ray binary makes certain that the compact source is 
a rotating neutron star.
For example, the transient X-ray binary V0332+53 was 
identified as a neutron star after detection of 
$\sim$4.4-s periodic coherent pulsations -- 
previously the source had been considered as a 
black hole candidate because the PDS of the source closely resembles those 
of the canonical black hole candidate Cyg X--1
(Lazzatti \& Stella 1997, and references therein).

Periodic and aperiodic variabilities of AXBPs 
have usually been studied independently.
Recent studies, however, suggest the possibility of coupling 
between them (Lazzatti \& Stella 1997; Burderi et al. 1997), and
Moon \& Eikenberry (2001) presented an observational evidence 
in large X-ray flares from the AXBP LMC X--4.
It is still not clear, however, exactly when and how the coupling happens.
The purpose of this paper is to present another observational
example of the coupling in the AXBP Her X--1,
together with the discovery of $\sim$12-mHz X-ray QPOs of the source.

Her X--1 is an eclipsing AXBP with pulsational and orbital periods of
$\sim$1.24 s and $\sim$1.7 d, respectively.
Its optical companion is a normal star of $\sim$2 $M_{\odot}$,
making Her X--1 a rare X-ray binary 
which is not readily classified as
a high-mass X-ray binary or a low-mass X-ray binary.
Her X--1 has shown very complex temporal features:
First, it is one of a few AXBPs showing a 
superorbital period ($\sim$35 d), 
usually ascribed to the motion of a precessing, warped accretion disk
periodically obscuring the observer's line-of-sight to the central neutron star.
The $\sim$35-d period consists of at least three sub-states:
(1) $\sim$7 orbital period main-on state, 
(2) $\sim$5 orbital period short-on state, and 
(3) two $\sim$4 orbital period low states between (1) and (2).
Secondly, Her X--1 has shown two different types of dips: 
pre-eclipse dips and anomalous dips,
due to obscuration by the surrounding accretion disk.
Thirdly, Her X--1 sometimes shows anomalous low-states 
(e.g., Coburn et al. 2000).
In addition,  Her X--1 was recently found to show mHz QPOs 
in its ultraviolet (UV) continuum, 
probably emitted on the surface of the  companion by reprocessing
the X-rays from the neutron star (Boroson et al. 2000).

\section{Observations and Analysis}

The main-on state of Her X--1 was observed with 
the {\it Rossi X-Ray Timing Explorer} (RXTE)
on 1996 July 26 and 27 during $\sim$27 hours of 
observation of the source.
The photon arrival times from the Good Xenon data
of the Proportional Counter Array were transformed 
to the solar system barycenter
using the JPL DE400 ephemeris.
Sixteen data segments, each with lengths of 0.5--1 h,
including five segments of pre-eclipse dips were obtained.

\subsection{Pre-Eclipse Dips}
%

Two lightcurves of the pre-eclipse dips, 
obtained with the five RXTE Proportional Counter Units 
in 2--30 keV energy range,
are shown in Fig. 1 with 4-s time resolution.
The strong aperiodic variabilities are clearly identified in Fig. 1, 
and we shall call the two lightcurves ``PED 1" and ``PED 2", respectively.
The phase of the $\sim$35-d super-orbital motion, 
based on the ephemeris of Scott \& Leahy (1999),
is $\sim$0.125, while the $\sim$1.7-d binary orbital phases,
based on the ephemeris of Deeter et al. (1991),
of PED 1 and PED 2 are $\sim$0.84 and $\sim$0.88, respectively.
The Leahy-normalized PDS of the lightcurves obtained with 2$^{-6}$ s 
time resolution are presented in Figure 2.
Only the first $\sim$0.71-h data of PED 1 were used in order to avoid data gaps,
while the full $\sim$0.92-h data were used for PED 2.
Fig. 2 shows the PDS around the coherent pulse frequency ($\sim$0.81 Hz) in the main window, 
the low-frequency (mHz) PDS in the small windows on the left side (in the logarithmic frequency scale),
and the detailed structures of the sidebands around the coherent frequency 
in the small windows on the right side.
Three characteristics are evident in the PDS of Fig. 2: 
(1) a strong peak at the coherent pulse frequency, 
(2) well-developed significant sidebands around the coherent frequency, 
and (3) strong low-frequency variability at $\sim$1--10 mHz.
The rms fractions
of the low-frequency components are
$\sim$30 $\pm$ 5 \% and $\sim$64 $\pm$ 10 \% for 
PED 1 and PED 2, respectively,
while those of the coherent pulsations 
($\nu$ = 0.795--0.820 Hz including the sidebands) are
$\sim$47 $\pm$ 5 \% for both PED 1 and PED 2.
%

The mHz powers and the sidebands around the coherent
frequency in Fig. 2 are compared in Fig 3,
where the mHz powers are overlaid onto the sidebands as follows:
First, the mHz powers are shifted to the frequency of the
higher-frequency (HF) sidebands, and then a lower-frequency (LF) mirror image 
was made with respect to the coherent frequency.
Finally, the shifted mHz powers are scaled to match the sidebands.
In Fig. 3, very similar distributions between the shifted, 
scaled mHz powers (crosses) and the sidebands (solid histogram) are identified, 
indicating that the amplitudes of periodic coherent pulsations 
are modulated by those of the aperiodic variabilities (or vice versa).
Linear correlation coefficients between them
are 0.880 $\pm$ 0.003 and 0.982 $\pm$ 0.001 over 74 and 78
bins for PED 1 and PED 2, respectively.
We estimated the uncertainties in the correlation coefficients
using a Monte Carlo simulation, creating 1000 simulated sideband/mHz power
distributions. Each simulated distribution had 74 (for PED 1; 78 for PED 2) 
Fourier bins drawn with replacement from the actual sideband/mHz power distributions.  
The quoted uncertainties are the maximum deviations from the actual correlation
coefficients over these 1000 simulations, and thus correspond approximately
to 99.9\% confidence level.  
We performed an additional Monte Carlo simulation as a null test, 
scrambling the order of the mHz powers while
maintaining the order of the sideband powers.  This resulted in average
correlation coefficients of 0.007 and -0.001 respectively for PED 1 and
PED 2, with standard deviations of 0.114 in each case, confirming that
the observed correlations are statistically significant.
%

\subsection{Normal State}

We show the lightcurve, PDS, and comparison between
the shifted, scaled mHz powers and the sidebands around the pulse
frequency of a $\sim$0.7-h gapless data segment obtained
during the normal state in Figure 4. 
Initial orbital phase of the data segment is $\sim$0.22.
The strong aperiodic variabilities (and the corresponding large mHz powers), 
which have been seen in Fig. 1 and 2 during pre-eclipse dips, are absent in Fig 4;
however, there exists a relatively weak (compared to the pre-eclipse dips)
correlation between the shifted, scaled mHz powers and the sidebands in Figure 4(c).
A linear correlation coefficient, obtained by fitting 62 bins,
is 0.492 $\pm$ 0.020, which is much smaller than those of the pre-eclipse dips
but still indicates a statistically significant (i.e., $\sim$95 \% confidence level) correlation.
A linear correlation coefficient between randomly sampled mHz powers and sidebands
are --0.002 $\pm$ 0.126.
%

Fig. 5 shows the averaged PDS (solid histogram) of the normal state
obtained by averaging the PDS of 13 gapless data segments of 1200-s length. 
The most conspicuous feature in Fig. 5 is the increase of the power intensity
at the low-frequency range (i.e., red noise)
with an excess component at $\nu$ $\simeq$ 12 mHz.
The PDS at $\nu$ $<$ 0.03 Hz were analyzed by $\chi^2$-fitting in two different ways. 
First, a power-law component and a Lorentzian component 
were assumed for the red noise and the excess component in the fitting.
Next, instead of the power-law component,
a Gaussian component (centered at $\nu$ = 0 Hz) and a Lorentzian component
were used. In both fittings, a background power intensity 38 was assumed,
which is evident from the PDS at higher frequency.
While the reduced $\chi^2$ from the first method is $\sim$2.5, 
the value from the second method is $\sim$0.98, both with 15 degree of freedom.
It is worth, at this point, to note that the red noise in Fig. 5 does not seem to 
follow a power-law distribution, which has been frequently assumed as the red noise distribution 
in the literature.
The PDS in Fig. 5 in fact seems 
to indicate the existence of another component at $\nu$ $\simeq$ 5 mHz. 
Although another component might exist, 
the short time duration (i.e., poor frequency resolution) of our data segments
makes it impossible to resolve the component.
The central frequency and full width half maximum (FWHM) of the Lorentzian component 
obtained with the second method are 12.4 $\pm$ 2.2 mHz and  
4.9 $\pm$ 3.5 mHz, respectively. 
The uncertainties correspond to $\sim$68.3 \% confidence level.
The results of the fitting is presented in Fig. 5 (dotted histogram).
The symmetric distribution and relative width (FWHM/central frequency)
of the excess component seem to satisfy the condition of being 
acknowledged as QPOs (van der Klis 1995).
A signal to noise ratio of $\sim$3.3 was obtained for the QPOs in Fig. 5
using the method descrbied by Boirin et al. (2000).
The rms fraction of the QPOs was calculated to be $\sim$1.7 $\pm$ 0.6 \% 
over the FWHM.
The central frequency, width, and rms fraction of the QPOs are roughly
comparable to the $\sim$8 $\pm$ 2 mHz UV QPOs (Boroson et al. 2000). 
%

\section{Discussion} 

\subsection{Coupling between Periodic and Aperiodic Variability}

It is reasonable to attribute the origin of the strong aperiodic variabilities
in the pre-eclipse dips to the variation of obscuration due to an inhomogeneous 
distribution of the absorbing material in the accretion disk,
integrated along the observer's line-of-sight to the central pulsar.
This, together with the very strong correlations between the 
mHz powers and the sidebands,
imply that both the amplitudes of the aperiodic variabilities
and the amplitude changes of the periodic coherent pulsations 
are proportional to the same parameter: 
the amount of obscuration of the X-ray beam from the pulsar by 
the accretion disk during the pre-eclipse dips.

Although most of previous studies have treated the periodic and aperiodic
variabilities independently, we might expect
the coupling between them because most of the aperiodic variabilities
are due to the time-varying funneling of the accreting columns onto 
the magnetic poles of the pulsars and, subsequently, the aperiodic variabilities
should be modulated by the pulsar rotations.
Recently, Lazzati \& Stella (1997) and Burderi et al. (1997) 
have demonstrated models with correlations 
between the PDS broadening of the wings 
around coherent peaks and the red-noise components, 
indicating couplings between the periodic and aperiodic variabilities.
The results also cast doubt on the reported
apparent correlation between the pulse frequency and the knee frquency,
below which the PDS steepens, of X-ray pulsars (Lazzati \& Stella).
The correlation between the mHz powers and sidebands found
in the normal state of Her X--1 in this study seems to be consistent with 
the results (Lazzati \& Stella 1997; Burderi et al. 1997),
which predict the origin of the aperiodic variabilities (i.e., red noise)
to be the accretion flows near the stellar surface.

\subsection{The X-ray QPOs}

The global-normal disk oscillations model (GDOM),
which is recently proposed by Titarchuk \& Osherovich (2000), 
seems to be capable of accounting for the $\sim$12-mHz X-ray QPOs.
In GDOM, $\sim$0.01--1 Hz persistent oscillations 
reported in several X-ray binaries
can be produced in the accretion disks surrounding the compact X-ray sources
as a result of the gravitational interaction between the
central sources and the accretion disks.
The oscillation frequency is a function of several parameters including
inner-disk radius ($R_{\rm in}$), outer-disk radius ($R_{\rm out}$), 
adjustment radius ($R_{\rm adj}$, below which surface density is assumed to be constant),
power index of the surface density distribution ($\gamma$),
and mass of the central source ($M_{\rm X}$).
The expected inner-disk radius of Her X--1 responsible for the X-ray QPOs 
is given by
\begin{equation}
R_{\rm in} \simeq 3.1 \times 10^5 \; \nu_{\rm QPO}^{-5/4} \; \; \rm cm
\end{equation}
where $\nu_{\rm QPO}$ is in unit of Hz. 
We have used $R_{\rm out}$ = 1.7 $\times $ 10$^{11}$ cm (Howarth \& Wilson 1983),
$R_{\rm adj}$ = 3 $R_{\rm in}$ and $\gamma$ = 3/5 (Titarchuk \& Osherovich), 
and $M_{\rm X}$ = 1.4 $M_{\odot}$.
The resulting inner-disk radius corresponding to the $\sim$12 mHz QPOs
is $R_{\rm in}$ $\simeq$ 0.5--1.3 $\times$ 10$^8$ cm.

For the case of AXBPs with strong magnetic field strength ($B$ $\sim$ 10$^{12}$ G),
such as Her X--1, the magnetic field is expected to disrupt 
the inner accretion disk around the 
magnetospheric radius, $r_{\rm M}$ $\simeq$ $\xi$$r_{\rm A}$,
where $\xi$ is a dimensionless parameter of $\le$ 1
and $r_{\rm A}$ is the Alf$\rm \acute v$en radius at which magnetic pressure is in
equilibrium with gas pressure
(Shapiro \& Teukolsky 1983; Bildsten 1997).
Using X-ray luminosity $L_{\rm X}$ $\simeq$ 2.1 $\times$ 10$^{37}$ ergs s$^{-1}$ 
(Choi et al. 1994), the assumed neutron star radius $R$ = 1 $\times$ 10$^6$ cm,
and the magnetic field strength $B$ $\simeq$ 3 $\times$ 10$^{12}$ G 
estimated from the analysis of cyclotron resonance 
features (Makishima et al. 1999, and references therein),
we obtain $r_{\rm A}$ $\simeq$ 5.4 $\times$ 10$^8$ cm for an 1.4 $M_{\odot}$ neutron star.
If we consider various assumptions used in obtaining the Alf$\rm \acute v$en
radius (e.g., Shapiro \& Teukolsky) and in the GDOM, 
the expected magnetospheric radius seems to be roughly comparable to the 
inner-disk radius of the GDOM calculated above.
(We note that Titarchuk \& Osherovich [2000] accounted for the 48 $\pm$ 2 mHz UV QPOs
of Her X--1 by assuming $R_{\rm in}$ $\simeq$ 1 $\times$ 10$^7$ cm,
which is much smaller than the magnetospheric radius.
It was not explained, however, how the inner disk can penetrate the 
strong magnetospheric barrier of Her X--1 to near the surface 
of the central pulsar.) 

On the other hand, Boroson et al. (2000) discussed in detail the 
8 $\pm$ 2 mHz UV QPOs from Her X--1 in terms of two canonical models for QPOs,
the beat-frequency model and the Keplerian-frequency model.
According to them, the UV QPO frequency,
in principle, 
can be explained by either the BFM or the KFM.
Since the frequency is comparable to the frequency of 
the normal-state X-ray QPOs of this study,
we ascribe the detailed discussion of the 
origin of the normal-state QPOs via these models to Boroson et al. (2000).

\section{Conclusions} 

The results of this paper are summarized as follows:
\begin{itemize}
\item
We have discovered coupling between the periodic
aperiodic variabilities from the AXBP Her X--1
when the source is in the pre-eclipse dip and normal state, respectively.
The coupling suggests that the amplitudes of 
the periodic coherent pulsations are modulated by the amplitudes of the 
aperiodic variabilities.
\item
The coupling in pre-eclipse dips indicates that both
the periodic and aperiodic variabilities are related to the same parameter:
the amount of obscuration of X-ray beam from the pulsar by the accretion
disk during the pre-eclipse dips.
\item
The normal state coupling may indicate that
the aperiodic variabilities from X-ray binary pulsars are due to 
time-varying accretion flows onto the pulsar's magnetic poles.
\item
We have also discovered $\sim$12-mHz X-ray QPOs 
which are comparable to the previously detected UV QPOs from the source.
\item
If the $\sim$12-mHz X-ray QPOs are due to the global-normal
disk oscillations caused by the gravitational interactions between the central pulsar
and the accretion disk, the inferred inner-disk radius
is roughly comparable to the magnetospheric radius,
$\sim$ 1 $\times$10$^8$ cm.
\item
The results of this paper, 
together with our previous results on LMC X--4 (Moon \& Eikenberry 2001),
offer unique opportunities for studying timing behaviour of 
accretion-powered X-ray binary pulsars.
\end{itemize}

\acknowledgments
We would like to thank the anonymous referee for comments
and suggestions which substantially improved this paper.
DSM acknowledges Wynn Ho for his careful reading of manuscript
and comments.
This research has made use of data obtained from the {\it
High Energy Astrophysics Science Archive Research Center}
(HEASARC), provided by NASA's Goddard Space Flight Center.
DSM is supported by NSF grant AST-9986898.
SSE is supported in part by an NSF Faculty Early Careeer Development
(CAREER) award (NSF-9983830).

\clearpage



\clearpage
\begin{figure}
\plotone{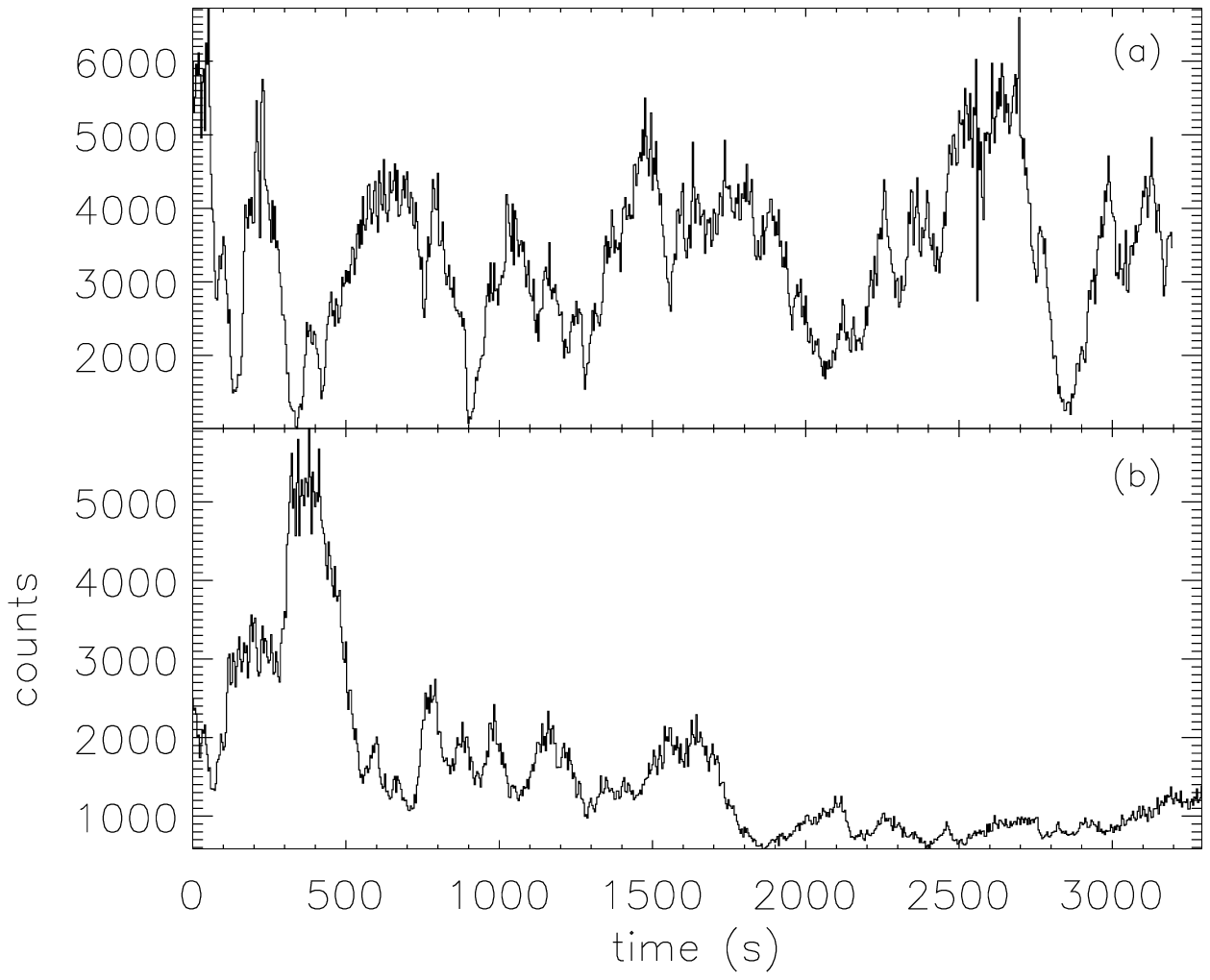}
\caption{Lightcurves, in 2--30 keV energy range with 4-s time resolution,
of Her X--1 during pre-eclipse dips : (a) for PED 1 and (b) for PED 2.
\label{fig1}}
\end{figure}

\clearpage
\begin{figure}
\plotone{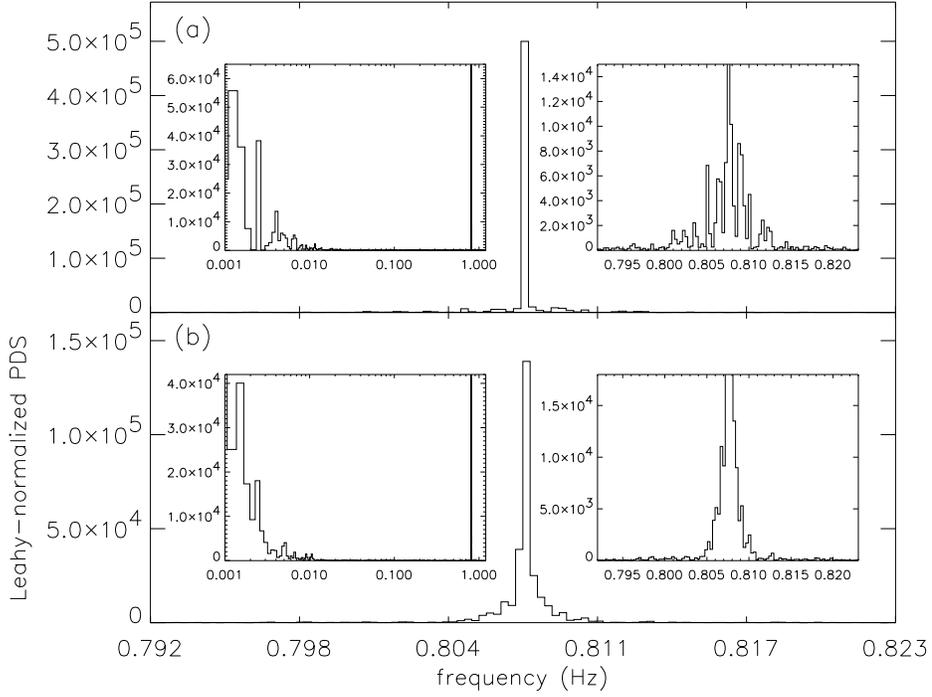}
\caption{Power-density spectra (PDS) of the lightcurves in Fig. 1:
(a) PED 1 and (b) PED 2.
The main windows show the PDS around the coherent pulse frequency.
The small windows on the left side show the mHz PDS
in the logarithmic frequency scale, while the small windows on the right side
show the detailed structures of the sidebands around the coherent frequency.
\label{fig2}}
\end{figure}

\clearpage
\begin{figure}
\plotone{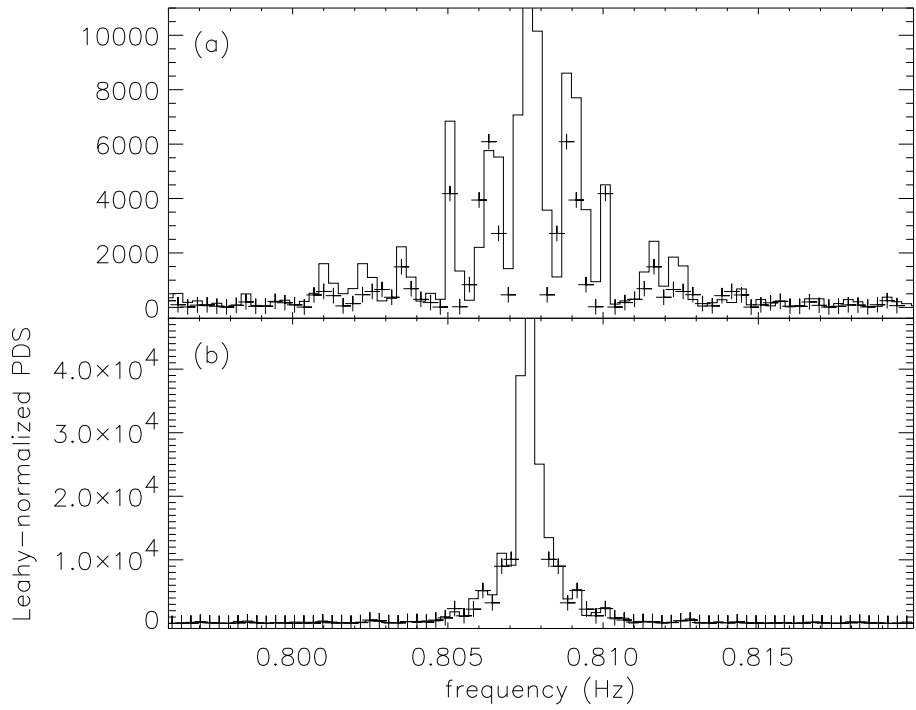}
\caption{(a) Comparison between the PDS of the sidebands (solid histogram)
and the shifted, scaled PDS (crosses) of the mHz
QPOs for (a) PED 1 and (b) PED 2.
\label{fig3}}
\end{figure}

\clearpage
\begin{figure}
\plotone{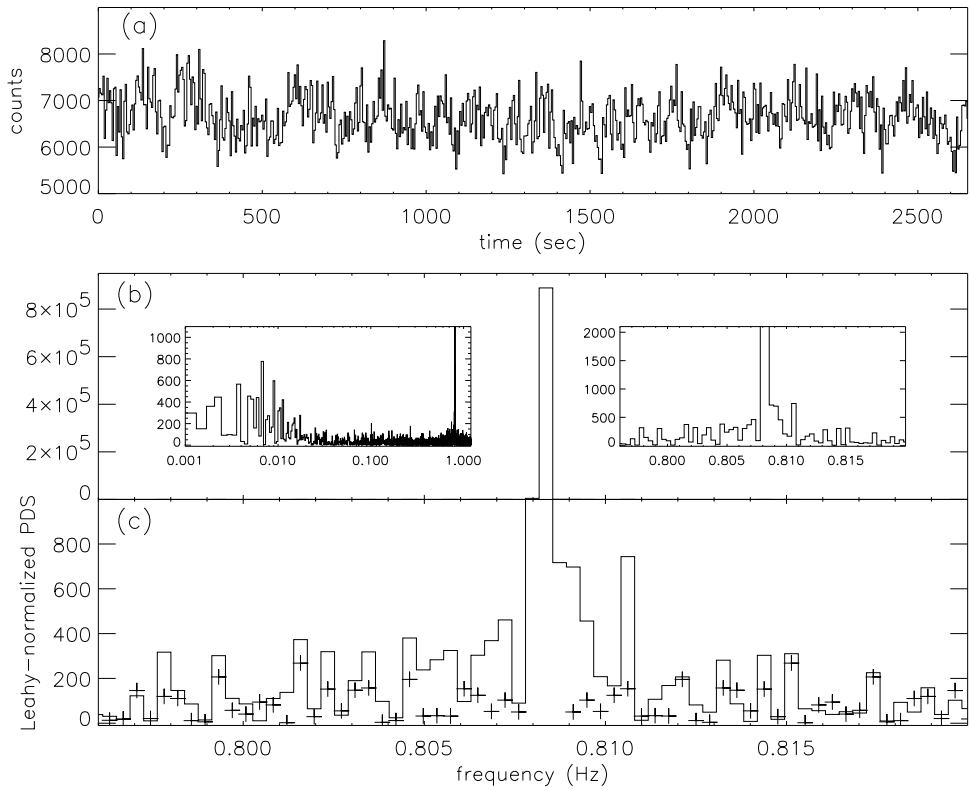}
\caption{(a) Same as Fig. 1, but for the $\sim$0.7-h gapless data segment 
of the normal state.
(b) Same as Fig. 2, but for the lightcurve of (a).
(c) Same as Fig. 3, but for the PDS of (b).
\label{fig4}}
\end{figure}

\clearpage
\begin{figure}
\plotone{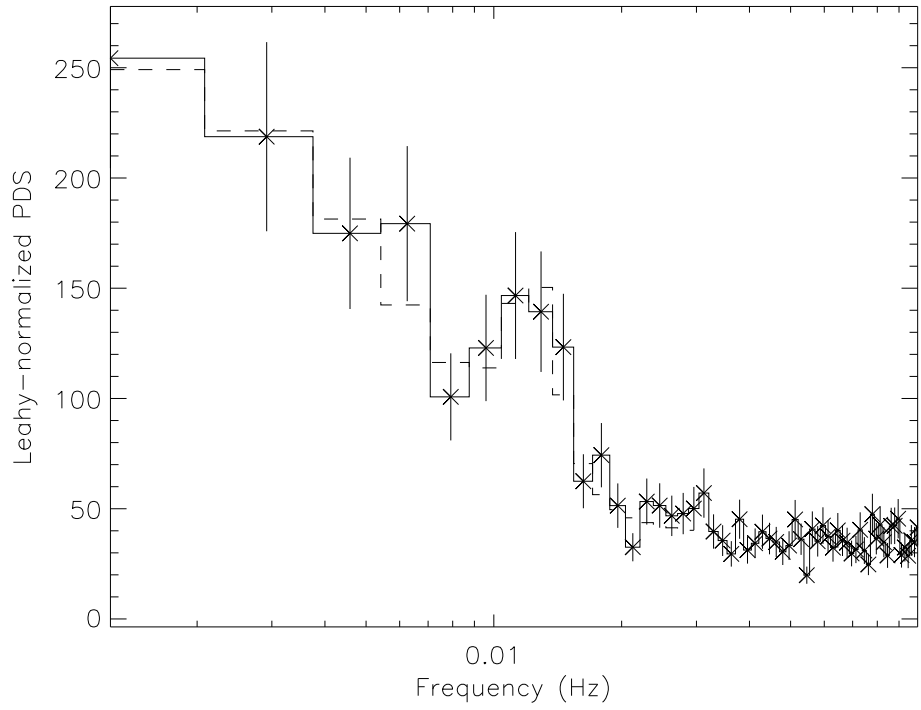}
\caption{Superposition of the PDS of the normal state (solid histogram)
obtained by averaging PDS of 13 data segments of 1200-s length
on the PDS obtained by fitting the averaged PDS with one
Gaussian component and one Lorentzian component (dotted histogram).
The error bars represent $\sim$68.3 \% confidence level.
\label{fig5}}
\end{figure}


\end{document}